\documentstyle[11pt]{article}
\normalbaselines
\begin{document}
\bibliographystyle{unsrt}

\def\bea*{\begin{eqnarray*}}
\def\eea*{\end{eqnarray*}}
\def\ba{\begin{array}}
\def\ea{\end{array}}
\count1=1
\def\be{\ifnum \count1=0 $$ \else \begin{equation}\fi}
\def\ee{\ifnum\count1=0 $$ \else \end{equation}\fi}
\def\ele(#1){\ifnum\count1=0 \eqno({\bf #1}) $$ \else \label{#1}\end{equation}\fi}
\def\req(#1){\ifnum\count1=0 {\bf #1}\else \ref{#1}\fi}
\def\bea(#1){\ifnum \count1=0   $$ \begin{array}{#1}
\else \begin{equation} \begin{array}{#1} \fi}
\def\eea{\ifnum \count1=0 \end{array} $$
\else  \end{array}\end{equation}\fi}
\def\elea(#1){\ifnum \count1=0 \end{array}\label{#1}\eqno({\bf #1}) $$
\else\end{array}\label{#1}\end{equation}\fi}
\def\cit(#1){
\ifnum\count1=0 {\bf #1} \cite{#1} \else 
\cite{#1}\fi}
\def\bibit(#1){\ifnum\count1=0 \bibitem{#1} [#1    ] \else \bibitem{#1}\fi}
\def\ds{\displaystyle}
\def\hb{\hfill\break}
\def\comment#1{\hb {***** {\em #1} *****}\hb }

\newcommand{\TZ}{\hbox{\bf T}}
\newcommand{\MZ}{\hbox{\bf M}}
\newcommand{\ZZ}{\hbox{\bf Z}}
\newcommand{\NZ}{\hbox{\bf N}}
\newcommand{\RZ}{\hbox{\bf R}}
\newcommand{\CZ}{\,\hbox{\bf C}}
\newcommand{\PZ}{\hbox{\bf P}}
\newcommand{\QZ}{\hbox{\bf Q}}
\newcommand{\HZ}{\hbox{\bf H}}
\newcommand{\EZ}{\hbox{\bf E}}
\newcommand{\GZ}{\,\hbox{\bf G}}
\vbox{\vspace{5mm}}

\begin{center}{\LARGE   
\bf Algebraic Geometry and Physics }\footnote{Talk
presented in The Third Asian Mathematical
Conference (AMC 2000) at Manila, Philippines,
October 2000.}
 \\[5mm]

{\bf Shi-shyr Roan  } 
\\[5mm] {\it Institute of
Mathematics \\ Academia Sinica \\  Taipei ,
Taiwan  \\ [3mm] ( Email:
maroan@ccvax.sinica.edu.tw )} \\ [15mm]

\end{center}

\begin{abstract}
This article is an interdisciplinary review and
an on-going progress report over the last few years
made by myself and collaborators in
certain fundamental subjects on two major
theoretic branches in mathematics and theoretical
physics: algebraic geometry and  quantum 
physics.  I shall take a practical approach,
concentrating more on explicit examples rather
than formal developments. Topics covered  are
divided in three sections: 
(I) Algebraic
geometry  on two-dimensional exactly solvable
statistical lattice models and its related
Hamiltonians: I will report results on
the algebraic geometry of rapidity curves
appeared in the chiral Potts model, and the
algebraic Bethe Ansatz equation in connection with 
quantum inverse scattering method for the related
one-dimensional Hamiltonion chain, e.g., XXZ,
Hofstadter type Hamiltonian.  
(II) Infinite
symmetry algebras arising from quantum spin chain
and conformal field theory: I will explain certain
progress made on Onsager algebra, the
relation with the superintegrable chiral Potts
quantum chain and problems on its
spectrum.  In conformal field theory,
mathematical aspects of characters of N=2
superconformal algebra  are discussed, 
especially  on the modular invariant 
property connected to the theory. 
(III).
Algebraic geometry problems on  orbifolds 
stemming from  string theory: I will report 
recent progress on crepant resolutions of
quotient singularity of dimension greater than or
equal to three.  The direction of present-day
research of engaging finite group
representations in the geometry of orbifolds is
briefly reviewed, and the mathematical aspect of
various formulas on the topology of string vacuum
will be discussed.

\end{abstract}

\section{Introduction}
In this report, I would like to summarize our 
recent work and on-going progress we have made in
the interdisciplinary area  of algebraic geometry
and  quantum  physics. A brief  historic account
relevant to the subjects will also be  presented. We
will discuss some fundamental problems in 
solvable theories in these fields. Here  the term
"solvable" is loosely used in the following sense:
in physics, it means some physical  quantities of
the  theory, e.g., the partition function, can be
computed exactly in a closed form; in mathematics,
it involves certain mathematical formulae or
equations expressed in an explicit form whose
solution can be effectively calculated. The
physical models  we are concerned with here are 
those among integrable statistical systems or 
string theory, where techniques or problems in
algebraic geometry naturally arise, in particular,
those related to 2-dimensional exactly solvable
statistical lattice models and orbifold models of
superstring compactification. It is known that a
2-dimensional  statistical lattice model 
associates a 1-dimensional quantum Hamiltonian
chain, whose eigenvalues determine the partition
function of the statistical system. Incorporating
the  symmetries of a quantum chain could  serve as a
useful means of  gaining solutions of the
original model.  Accordingly, the representation
theory of symmetrical algebras was brought in
within the context of solvable models. Recently
many such theories  have been explored along this
scheme in literature with spectacular progress.
But, just to keep things simple, we will in this
article restrict our attention only to a few such
kind of models we have been involved with at the
present day.  Let us address the main topics in
this report with a  comparison list of theories
involved in the areas of physics and mathematics:
$$
\begin{array}{ll}
\ \ \ \ \ {\bf Physics }   & \ \ \ \ \ {\bf
Mathematics }
\vspace{.2 cm}  \\ {\rm
(Integrable) \  Chiral \ Potts 
\ Model},  & {\rm Algebraic \ Curve } ; 
\vspace{.1 cm} 
\\ 
{\rm Superintegrable \ Chiral \ Potts \ Model}, &
{\rm Onsager \ Algebra \ }, sl_2{\rm-loop
\ Algebra} ; \vspace{.1 cm} \\
{\rm Hofstadter \ Type \ Hamiltonian}, & {\rm
Bethe  \ Equation \ on \ Riemann \ Surface }
; \vspace{.1 cm}
\\ 
{\rm N=2 \ Conformal \ Field \ Theory}, & {\rm
Elliptic \ Theta \ Form }; \vspace{.1 cm} 
\\
{\rm Orbifold \ Model \ in \ Superstring \ Theory
}, & {\rm Crepant \ Resolution }, \
G{\rm-Hilbert \ Scheme } \ .
\end{array} 
$$ 
Here I will summarize the research that we
undertook on these issue over the past five years.
The presentation is divided in three sections with
the references based upon our work in the
area:
\begin{itemize}
\item Exactly solvable integrable model: 
\cite{LR, R98}. 
\item Symmetry algebra of quantum spin
chain, superconformal field theory: 
\cite{DR, GRo, KRW,  R99}.
\item Orbifold, finite group representation
 and string theory:  \cite{CR, Rtop, R97}
\end{itemize}
{\bf Convention .} In this article, 
$\RZ, \CZ$ will denote 
the field of real, complex numbers respectively; 
$\ZZ$ the ring of integers.  For positive integers 
$n$ and $N$, we will denote
$\stackrel{n}{\otimes} \CZ^N$ the tensor
product of $n$-copies of the complex $N$-dimensional
vector space $\CZ^N$, and $\ZZ_N$ the quotient ring
$\ZZ/N\ZZ$.

\section{Exactly Solvable Integrable Model}

In this section, we discuss a class of
2-dimensional lattice models, called solvable
integrable models, in which certain interesting
physical quantities, in particular the partition
function, can be explicitly computed. The models
are built by assigning spins $\sigma_i$ on sites of
a (finite or infinite) planar lattice with adjacent
spins interacting along the edges by a rule of 
Boltzmann weights. The spins
$\sigma_i$ take values of $\pm 1$; or $N$
different values, $0,
\ldots, N-1 \in \ZZ_N$. 
For a such system, the Yang-Baxter (or
star-triangle) relation of Boltzmann weights
plays a central role in the theory of an exactly
solvable model in statistical mechanics and field
theory. For statistical lattice models, it is well
known that a model is solvable by techniques of 
commuting transfer matrices  if the
Boltzmann weights satisfy the Yang-Baxter equation.
There are a number of notable Yang-Baxter solvable
lattice models which have been extensively studied
in both literature of physics and
mathematics (see, e.g., \cite{B82}).
Here I will consider only a few of such kind of
models in which there exists  a certain intricate
relation with problems in algebraic geometry.

In the 1980s, a two-dimensional lattice integrable
model with phenomenological origin in physics,
known as the chiral Potts $N$-state model,
was found in statistical mechanics where the
"rapidity" (or "spectral variable") lies on a
algebraic curve of genus higher than one, ( A brief
history account can be found in Sect. 4.1 of  
\cite{Mc}). The
model takes 
$N (\geq 3)$ values for the spin per site, which is
a generalization of the renowned Ising
model first solved by L. Onsager in 1944
\cite{O}. The Boltzmann weights,
$W_{p,q},\overline{W}_{p,q}$,  have the form, 
\begin{eqnarray*}
\frac{W_{p,q}(n)}{W_{p,q}(0)}  = \prod_{j=1}^n
\frac{d_pb_q-a_pc_q\omega^j}{b_pd_q-c_pa_q\omega^j}
\ , \ \ \ \ \
\frac{\overline{W}_{p,q}(n)}{\overline{W}_{p,q}(0)} 
= \prod_{j=1}^n
\frac{\omega a_pd_q-
d_pa_q\omega^j}{ c_pb_q- b_pc_q \omega^j}
\end{eqnarray*}
where $\omega= e^{\frac{2 \pi i}{N}}$, $n \in \ZZ_N
$, and $p, q$ are two rapidities represented by 
the ratio of four-vectors, $a:b:c:d$, satisfying the
following set of equations,
\begin{eqnarray}
ka^N + k'c^N = d^N , & kb^N + k'd^N = c^N ,
\nonumber 
\\ a^N + k'b^N = kd^N , & k'a^N + b^N =
kc^N . \ 
\label{rapidF}
\end{eqnarray}
Here  $k, k'$ are the real parameter with the
relation $k^2 + k'^2 = 1$. The periodicity of $n$
modulo $N$ of the Boltzmann weights  follows
from the above constraints of rapidities, which 
form a curve of genus $N^3-2N^2+1$ in the
projective $3$-space $\PZ^3$. For a lattice with
the horizontal size $L$ and 
periodic boundary condition, the combined weights of
intersection between two consecutive rows give rise
to the transfer matrix, 
\begin{eqnarray*}
T_{\{ \sigma, \sigma'\}}(p, q) = \prod_{l=1}^L
\overline{W}_{p,q}(\sigma_l - \sigma_l')
W_{p,q}(\sigma_l - \sigma_{l+1}') \ ,
\end{eqnarray*}
where $\sigma= (\sigma_1, \ldots, \sigma_L) $ ,
$\sigma'=(\sigma_1',
\ldots, \sigma_L') $ with $\sigma_l, \sigma_l' \in
\ZZ_N$. The transfer matrix $T(p, q)$ is an
operator acting on the  $L$-tensor vector space of
$\CZ^N$,
$\stackrel{L}{\otimes} \CZ^N$. Here the coordinates
of  a vector in 
$\CZ^N$ will be indexed by elements in $\ZZ_N$,
equivalently to say, it is represented by a
sequence of coordinates,
$ \{ v_k \}_{k \in \ZZ}$, with the $N$-periodic
condition, $v_k = v_{k+N}$. The complete 
spectrum of the operator $T(p, q)$ is 
one of the main mathematical 
problems of physical interest. By  the
star-triangle relation satisfied by  Boltzmann
weights, it ensures the commuting relation of
transfer matrices for a fixed rapidity $p$:
$$
[ T(p, q) , \ \ T(p, q') ] = 0 \ .
$$
Furthermore, $T(p, q)$ also satisfies a certain
functional equation involving an automorphism of
the Riemann surface defined by $R(a_q, b_q, c_q,
d_q) = (b_q,
\omega a_q, d_q, c_q)$. Note that for $N=2$, it is
the Ising case where (\ref{rapidF}) defines an
elliptic curve, known to carry a canonical additive
group structure. Solutions of the functional
equation can be obtained by using elliptic theta
function representations of  $a,b, c, d$
in the rapidity. However, the "difference property"
of rapidities no longer holds for the spectral curve
(\ref{rapidF}) as $N
\geq 3$. This creates the peculiar character which
makes the chiral Potts model  distinct with all
previously known models, hence inspired  the
algebro-geometry approach on studying 
integrable models with a such kind of nature. Note
that the curve (\ref{rapidF}) possesses a large
finite number of symmetries, and
the normalized Boltzmann weights are meromorphic
functions of that Riemann surface.  In various
problems of computating the physical quantities,
e.g., free energy, order parameter etc., the
spectral curve  can be reduced to some lower genus
one by modulating certain symmetry subgroup.  For
the spectrum of the transfer matrix,  the curves are
the ones obtained from (\ref{rapidF}) quotiented by
a certain order
$N^2$ symmetry subgroup, with the variables,
$t=
\frac{ab}{cd},
\lambda= (\frac{d}{c})^N$, satisfying the relation,
\begin{eqnarray}
{\cal C}_{N, k'} : \ \  t^N = 
\frac{(1-k' \lambda ) ( 1 - k'
\lambda^{-1} ) }{k^2} \ . \ \label{Chyp}
\end{eqnarray}
This is one-parameter curves with the  
characterization of the genus $N-1$ hyperelliptic
curves carrying the dihedral symmetrical group
$D_N$  structure. With the functional
equation derived in \cite{BBP, BazS}, the
computation  was made 
in the infinite lattice, (i.e.,
$L \rightarrow \infty$), to obtain a closed form 
for the "ground state" energy  in \cite{Bax90} and
the excitation energy in \cite{MR, Geh}, by an
ingenious method devised by R. J. Baxter bypassing
the algebraic geometry of the  hyperelliptic curves.
These have been the only few of progresses made on
the physical computations of spectrum to this day.
As the solutions involved are rational functions on
Riemann surfaces, which in principal could be
constructed from its zeros and poles, techniques
 in algebraic geometry would be expected
of great help to such  problems. 
Through the classical work of S. Kowalevski,   R.
J. Baxter first found  the expression of
hyperelliptic function parametrization of Boltzmann
weights years ago \cite{Bax}; subsequently
 the Baxter's parametrization was  explored in
\cite{R92} by an algebraic geometry  method where
the prime form expression of the  Boltzmann weights
was identified. In an recent work
\cite{R98},  differential equations for the periods
and the expression of the modulus of the
hyperellpitic family (\ref{Chyp}) in terms of the
Schwarz's triangle function were obtained. However
all these findings in algebraic geometry, though
interesting in a certain sense, haven't 
 yet provided any significant  insight in solving 
the original eigenvalue problem. Similar
difficulties occur on the role of algebraic geometry
 when applying to solutions of other
physical quantities, among which a notable one is
the strong conjecture made in 
\cite{AMPT} 
on the order parameter with a Ising-like form:
$$
\langle Z_0^j \rangle = ( 1-
k'^2)^{\frac{j(N-j)}{2N^2}}
\ , \ \ 
0 \leq j \leq N-1 \ . 
$$
After many unsuccessful attempts of employing
algebraic geometry into problems of
the chiral Potts model, B.M. McCoy made the
following epigram in \cite{Mc} which goes as
: 

\begin{em}
The nineteen century saw many
brilliant creations of the human mind. Among them
are algebraic geometry and Marxism. In the late
twentieth century Marxism has been shown to be
incapable of solving any practical problem but we
still do not know about algebraic geometry .
\end{em}

\par \vspace{.1cm} \noindent
Nevertheless, the study of chiral Potts model
does raise some of the most intractable
problems in algebraic geometry at the present day;
the quest of rigorous mathematical theorems with
 physical implication  should be 
an interesting, though difficult, program in
algebraic geometry. On the other hand, in the
physical content there are  some notable
1-dimensional Hamiltonian chains  which are
speculated to have certain  intimate connection
with the chiral Potts model;  though  the exact
formulation is not  apparently clear right now.
Among these Hamiltonions, the XXZ spin chain,
Hofstadter type Hamiltonian, discrete sine-Gordon
model  are included. It is well known that the XXZ
chain is the one corresponding to the  2-dimensional
six-vertex statistical model 
\cite{Lb, YY},  a 
renowned solvable system whose structure
was the primary source of the 
mathematical development of quantum
group theory (for a pedagogical account,  see e.g.
\cite{F92, Jim}). By  
the work of Bazhanov and Stroganov \cite{BazS}, it 
proposed that the chiral Potts model can be
regarded as a certain new algebraic structure
related to the six-vertex model. By this suggestion,
the quantum inverse method approach developed by the
Leningrad school in the early eighties \cite{TF}
could be brought into the study of those related
spin chains, which would possibly provide  some 
valuable insights to the chiral Potts model. Indeed
it is the case with certain discrete integrable
models -with discretized space and time
variables-involved. The  Hofstadter type Hamiltonian
is a such example with progress we recently made.

The Hofstadter Hamiltonian has the following form
for parameters $\mu, \nu \in \RZ$, 
\begin{eqnarray*}
H_{Hof}  =  \mu (    U +
 U^{-1})  + \nu (  V + V^{-1} )  ,
\end{eqnarray*}
where $U, V$ are unitary
operators satisfying the Weyl commutation relation:
$UV= \omega VU $, 
 $\omega := e^{2 \pi \sqrt{-1} \Phi}$ with the
phase $\Phi \in \RZ$. The Hamiltonion can be
interpreted as  a 2-dimensional Bloch system in
the presence of constant electric and magnetic
fields, which has been a model with a long and rich 
history in physics. The phase 
$\Phi$ represents the magnetic flux per plaquette.
This problem has attracted an unceasing interest in
both the physical and mathematical communities since
the 1976 numerical work of Hofstadter \cite{Hof}
exhibited a beautiful fractal structure of butterfly
spectral band versus the magnetic flux.
For a rational flux $\Phi$, a surprising connection
was recently found  between problems of
diagonalizing the Hofstadter type Hamiltonians
and solving the 1-dimensional spin chain
of a finite site by the quantum transfer
matrix  method. The development was
motivated by the work of Wiegmann and
Zabrodin \cite{WZ}
 on the appearance of quantum $U_q(sl_2)$
symmetry in magnetic translation problems. 
Subsequently in
\cite{FK}, Faddeev and Kashaev considered the
following Hamiltonian, including  the Hamiltonian 
$H_{Hof}$  as a certain special limit,
\be
H_{FK}  =  \mu (    U +
 U^{-1})  + \nu (  V + V^{-1} ) + \rho ( W +
 W^{-1}) ,
\ele(FKHam)
where $U, V, W$ satisfy the pair-wise Weyl
commutation relations: $
U V= \omega VU $, $ VW = \omega WV$, $ WU=
\omega UW$. For a rational flux $\Phi= P/N$ with
$\omega$ a primitive $N$-th root of unity, by the
well known result on representations of the Weyl
algebra, one can rewrite the operators $U,V, W$ in
$H_{FK}$ as $\alpha U, \beta V, \gamma W$ with
$\alpha, \beta, \gamma \in \CZ$, plus the
 $N$-power identity of $U, V, W$:
$U^N=V^N=W^N=1$. The
Hamiltonian $H_{FK}$ can be realized in the transfer
matrix of a Yang-Baxter model for the
six-vertex
$R$-matrix; its Yang-Baxter solution
${\cal L}$-operator, 
appeared also  in the study of chiral
Potts $N$-state model \cite{BBP}, has the
following form:
$$
{\cal L}_h (x) = \left( \begin{array}{cc}
       aZX  & xbX  \\
        xcZ &d    
\end{array} \right) \ , \ \ x \in \CZ \ ,
$$
where $X, Z$ are  operators of $\CZ^N$
defined by
\begin{eqnarray}
X, Z : \CZ^N \longrightarrow \CZ^N  , &
X|m \rangle := |m+1 \rangle , \ \ Z|m\rangle :=
\omega^m |m\rangle \ ,  \ \label{XZ}
\end{eqnarray}
where $m \in \ZZ_N $, and  the parameter $h$
represents the ratio of four complex numbers,
$a:b:c:d$, i.e., an element in $\PZ^3$. It follows
that the ${\cal L}$-operator of a size $L$ ( with 
operators of
$\stackrel{L}{\otimes}\CZ^N$ as its entries),
\begin{eqnarray*}
{\cal L}_{\vec{h}}(x) =
{\cal L}_{h_0}(x)
\otimes {\cal L}_{h_1}(x) 
\otimes \ldots \otimes {\cal L}_{ h_{L-1}}(x)  \ , 
\end{eqnarray*}
again satisfies the Yang-Baxter relation.  
The transfer matrices, 
$$
T_{\vec{h}} (x) := {\rm tr} ({\cal L}_{\vec{h}}(x)
, x \in \CZ ,
$$ 
form a family of commuting
operators of $\stackrel{L}{\otimes} \CZ^N$, hence
with the common eigenvectors. The exact
spectra of
$T_{\vec{h}} (x)$ will be the main problem we are
concerned with.  A powerful  algebraic
Bethe-ansatz-technique in the investigation of
this type of problems was introduced in \cite{FK}
as follows. By applying a gauge
transformation of the ${\cal L}$-operator at each
site $j$ by introducing a set of new variables
$\xi_j$s,
\begin{eqnarray*}
\widetilde{\cal L}_{h_j}(x, \xi_j, \xi_{j+1}) =
\left(
\begin{array}{lc} 1 &  \xi_j-1\\
1 &  \xi_j 
\end{array}\right)
{\cal L}_{h_j}(x) \left( \begin{array}{lc}
1 &  \xi_{j+1}-1\\
1 &  \xi_{j+1} 
\end{array}\right)^{-1}
\ , 
\end{eqnarray*}
where $0 \leq j \leq L-1 $ and $\xi_L := \xi_0 $
it remains the same for the transfer matrix, i.e., 
$T_{\vec{h}} (x) = {\rm tr} (\bigotimes_{j=0}^{L-1}
\widetilde{\cal L}_{h_j}(x,
\xi_j, \xi_{j+1}))$. However, there exists a
family of vectors $|p\rangle$  
in $\stackrel{L}{\otimes} \CZ^N$ with the parameter
$p$ in a "spectral curve" 
${\cal C}_{\vec{h}}$, which is defined in $(x,
\xi_0,
\ldots, \xi_{L-1})$-space by the
relations,
\begin{eqnarray*}
{\cal C}_{\vec{h}} : & \xi_j^N  =(-1)^N
\frac{\xi_{j+1}^Na_j^N - x^Nb_j^N}{\xi_{j+1}^N x^N
c_j^N - d_j^N } \ \ , &  j = 0, \ldots, L-1 \ ,
\end{eqnarray*} 
such that $T_{\vec{h}} (x)$ acts on those selected
vectors $|p
\rangle$ in a "geometrical" manner, 
\begin{eqnarray*}
T_{\vec{h}}(x) |p\rangle = |\tau_- p\rangle 
\Delta_-(p)  + |\tau_+ p\rangle\Delta_+(p) \ .
\end{eqnarray*} 
Here $\tau_\pm$ are automorphisms, and
$\Delta_\pm$  the algebraic functions of  
${\cal C}_{\vec{h}}$ defined by 
$$
\begin{array}{l}
\tau_\pm :  (x, \xi_0, \ldots, \xi_{L-1}) 
\mapsto (q^{\pm 1} x, 
q^{-1} \xi_0, \ldots, q^{-1} \xi_{L-1}), \ \ \  \
q:=
\omega^{\frac{1}{2}} \\
\Delta_-(x, \xi_0, \ldots, \xi_{L-1})  =
\prod_{j=0}^{L-1}( d_j-x
\xi_{j+1} c_j ) ,  \\ 
\Delta_+(x, \xi_0, \ldots, \xi_{L-1}) = 
\prod_{j=0}^{L-1} \frac{\xi_j
(a_jd_j-x^2b_jc_j)}{\xi_{j+1}a_j -xb_j} \  . 
\end{array}
$$
For a common eigenvector $\langle \varphi|$ of
$T_{\vec{h}}(x)$  with the
eigenvalue
$\Lambda(x) \in \CZ[x]$, the function
$Q(p)= \langle \varphi|p\rangle$ of ${\cal
C}_{\vec{h}}$  
 satisfies the
following relation, called the Bethe
equation (of the model), 
\bea(l)
\Lambda(x) Q(p)  = Q(\tau_-(p)) \Delta_-(p) 
+  Q(\tau_+(p)) \Delta_+(p) \ , \ \ {\rm for} \ p 
\in {\cal C}_{\vec{h}} \ .
\elea(Bethe)
The family of vectors $|p\rangle$ is called the
"Baxter vector" or "Baxter vacuum state" in
literature, which was
explained in \cite{LR} from the aspect of algebraic
geometry. As the Bethe solutions
$Q(p)$ of (\req(Bethe)) are rational functions of
the algebraic curve ${\cal C}_{\vec{h}}$, the
algebraic geometry of ${\cal C}_{\vec{h}}$ would
play a vital role in the solvability of the Bethe
equation in this scheme. The complexity of the
problem heavily depends on the size $L$ and
the function theory of the corresponding curve 
${\cal C}_{\vec{h}}$. For the diagonalization
of Hofstadter type Hamiltonion 
(\req(FKHam)), it is related to  the case $L=3$, the
simplest one with respect to the size. However
even in this situation, the general solution of
(\req(Bethe)) remains an unsolved problem due to
the high genus of the curve
${\cal C}_{\vec{h}}$ though a primitive qualitative
investigation was conducted in \cite{LR}.

When the curve ${\cal C}_{\vec{h}}$ degenerates into
a finite collection of rational curves for
certain special values of $\vec{h}$, algebraic
geometry of the spectral curve becomes a trivial one
. With a detailed analysis of the form of Baxter
vector and a certain "diagonalization" procedure of
eliminating the effects of variables
$\xi_j$,  the equation (\req(Bethe)) is reduced
to a finite number of polynomial
Bethe equations, which possess a form of 
"Strum-Liouville like" difference equation on the
complex plane. Even with a simple expression
of these rational Bethe equations, the
determination of its solutions
$\Lambda(x), Q(x)$  for an arbitrary given size $L$
still posts a difficult mathematical problem, from
both the quantitative and qualitative aspects. For
$L=3$,
 we made a thorough mathematical study
of the rational Bethe equations in \cite{LR}, and
derived its complete solutions. The results have
revealed some certain mathematical characters
common with other standard intergable models as the
XXX-chain by using the algebraic
Bethe-ansatz-technique; one can discuss the
"degeneracy" of  eigenstates of the transfer matrix
with respect to the Bethe solutions. However,
inspired by the butterfly fractal structure of the
Hofstadter Hamiltonian, we adopt the infinity
procedure by letting $N \rightarrow \infty$ while
keeping the size $L$ fixed as the "thermodynamic"
limit, a consideration corresponding to the
irrational flux in the Block problem. We have made
the detailed discussion of these structures  with
possibly physical interest. It is worth noting that
the results obtained for rational Bethe solutions
are corresponding to those in the spectrum problem 
of the Hamilition (\req(FKHam))  with a special
choice of representations of Weyl operators $U,V,W$.
For a higher size
$L$, some mathematical  subtleties involved in
solutions of the rational Bethe equation are
difficulty to overcome at this moment. However, we
have now conducted a similar analysis on the case
$L=4$, due to its 
connection with  discrete quantum pendulum,
discrete sine-Gordon model. The progress we now
obtain has indicated a certain new feature unseen in
the case $L=3$. A deeper study would enhance
our understanding of the efficiency of this
algebraic method on diagonalizing integrable
Hamilitonians. But ultimately, the quest of possible
interplay of these analysis with the chiral Potts
model would be our primary concern for the
programme  undertaken by now.

\section{Symmetry Algebra of Quantum Spin
Chain, Superconformal Field Theory } 
The principle of symmetries has been one of the
main ideas of modern physics. If a model possesses
certain symmetries, its analysis becomes much
simpler by employing the representation theory of
the symmetrical algebra. In this section, we
present two such models which we have made certain
progress in recent years.

In the family of the chiral Potts $N$-state model,
 a  peculiar integrable structure occurs to enhance
on solving the model when the vertical rapidity
$p$ takes a special value. This special case
is called "superintegrable", which means the
vertical rapidity satisfies the relations:
$a_p=b_p  ,  c_p=d_p $.
The corresponding element in 
${\cal C}_{N, k'}$ of (\ref{Chyp}) has the
coordinate
$\lambda_p =1$, i.e., $p$ lies above a branched
point of the hyperelliptic curve ${\cal C}_{N,
k'}$. In this way,  the superintegrable case  can
be regarded as a general chiral Potts model with a
special symmetry condition on the rapidity. It turns
out the integrabilty of the model has  a stronger
sense than the usual ones. In this case, the
corresponding 
$N$-state  quantum chain  has the 
following  Hamiltonian form, 
\bea(c)
H(k') = H_0 + k' H_1 \ , \\  H_0 := -2
\sum_{l=1}^L
\sum_{n=1}^{N-1} \frac{X_l^n}{1-\omega^{-n}} ,
\ \ \  
 H_1 := -2 \sum_{l=1}^L \sum_{n=1}^{N-1} 
\frac{Z_l^nZ_{l+1}^{N-n}}{1-\omega^{-n}},
\elea(SIHam)
where  $X_l, Z_l$ are the operators on
$\stackrel{L}{\bigotimes}\CZ^N$ with the Weyl
operators
$X,Z$ of (\ref{XZ}) at the $l$-th site:   
$
X_l = I \otimes \ldots \otimes \stackrel{l{\rm
th}}{ X}
\otimes \ldots \otimes I $, $ 
Z_l = I \otimes \ldots \otimes \stackrel{l {\rm
th}}{ Z}
\otimes \ldots \otimes I ,  (Z_{L+1}= Z_1 ) $.
Note that the operator
$H(k')$ is  Hermitian for real $k'$, hence
with the real eigenvalues. 
When $N=2$,  $H(k')$ is 
the Ising quantum chain ; for $N=3$, one obtains
the 
$\ZZ_3$-symmetrical self-dual chiral clock model 
with the chiral angles $\varphi= \phi =
\frac{\pi }{2}$  in  \cite{HKN}. 
For a general $N$,
$H(k')$ was first constructed in a paper of 
G. von Gehlen and R. Rittenberg \cite{GR}, in which 
 the following Dolan-Grady 
condition \cite{DG} was shown to hold for the
operators
$A_0, A_1$ with
$A_0 = -2 N^{-1} H_0 , A_1
= -2 N^{-1} H_1 $, 
$$
[A_1, [A_1, [A_1, A_0]]]= 16 [A_1, A_0] \ , \ \ 
[A_0, [A_0, [A_0, A_1]]]= 16 [A_0, A_1] \ .
$$
The above relation is equivalent to  
the Onsager algebra \cite{Dav}, which is composed
of operators $A_m, G_m \ (m \in \ZZ) $ through the
recurrent relation via $G_1 :=  \frac{1}{4}[A_1,
A_0 ]$,
\begin{eqnarray*}
A_{m-1} - A_{m+1} = \frac{1}{2} [A_m , G_1 ] \ , & 
G_m = \frac{1}{4} [A_m, A_0 ] \ .
\end{eqnarray*}
This  Lie algebra satsifies  
following relations 
\begin{eqnarray*}
 [A_m , A_l ]  = 4 G_{m-l} \ , &
 [A_m , G_l ] = 2 (A_{m-l} - A_{m+l} ) \ ,
& [G_m, G_l] = 0 \ ,
\end{eqnarray*}
which appeared in the seminal paper \cite{O} of
L. Onsager in 1944  on the solution of
two-dimensional Ising model. Ever since then, an
intimate relationship has been known to exist
between Onsager algebra  and
$sl_2$, the exact connection was
clarified in \cite{ROns} through a mathematical
formulation by regarding the Onsager
algebra as the fixed subalgebra of 
$sl_2$-loop algebra, $sl_2[t,
t^{-1}]$,  by the canonical  involution. In a recent
joint work  with E. Date
\cite{DR}, we made a systematic study of the
algebraic structure of the Onsager algebra; 
obtained the complete classification of
"closed"
Lie-ideals\footnote{ Here a Lie-ideal $I$ of
a Lie algebra $L$ is called closed if
$L/I$ has the trivial center.} and the structure of quotient 
algebras. 
The key ingredient is to employ the
$sl_2$-loop variable $t$ by working on the
Laurent polynomials on the relevant problems which
we concern. The ideal theory of the Onsager algebra
arisen from the finite dimension representation of
the algebra has revealed a simple and elegant
structure, which provides a clear link between 
representations of Onsager algebra and 
$sl_2$-loop algebra.  With the
$t$-evaluation, the finite dimensional unitary
representations of Onsager algebra are derived
in a canonical manner.  Meanwhile, the essence
differentiating Onsager algebra from
$sl_2[t,t^{-1}]$ has indicated the
structure of certain solvable algebras and
a combinatorial nature of the content, though the
full explanation is still lacking by now. The
combinatoric implication of Onsager algebra would be
expected by recent progress made on subjects
related to a deformed Dolan-Grady relation
(see, e.g.,
\cite{Terw}). However, the mathematical analysis on
Onsager algebra  only ensures a certain special
Ising-like form for the  eigenavlues of the
Hamiltonian (\req(SIHam)), the precise expression
could not yet be derived  even for the 
"ground-state" energy, known in physical literature 
\cite{AMP, B88}. The expression involves the
Chebychev
 type polynomials for $N$=2
\cite{B82}. For
 higher $N$ ,  by a recent study on specific cases,
we are able to achieve the equations
satisfied by the Baxter-Albertini-McCoy-Perk
polynomials, and understand the nature and the
distribution of their zeros \cite{GRo}.
Nevertheless, the complete knowledge of the
spectrum of (\req(SIHam)) still remains a
theoretical mathematical challenge in the context
of study of Onsager algebra. 

Another type of infinite dimensional 
algebras  we  have been working on is the
superconformal algebra, which describes the
underling symmetries of superstring theory.  It is
known in physics that certain conformal field
theories can be identified with the continuum
scaling limit of some critical integrable
lattice models, a fact
which is served as a link between 2-dimensional  
statistical mechanics and string theory.
The supersymmetric extensions of spacetime
symmetries can  enhance profound structures in
physical content; in fact, supersymmetry is more or
less mandatory in string theory 
in which quantities of physical interest could be
calculated exactly. Though the subject
has been studied in physical literature extensively
since long (see 
\cite{DKRY} and references therein ), the
systematic  mathematical study of conformal
superalgebras has been taken up by mathematicians
only until recent years \cite{CKF}.   
Mathematically, the superconformal fields
are described by  a simple Lie  superalgebra,
called  a superconformal algebra. It is spanned
by modes of  a finite family of local fields,
containing the Virasoro and some other even,
odd ones, such that coefficients  of the operator
product expansions are linear combinations of
fields  in the family and their derivatives. Among
the many conformal field theories, the N=2
conformal theory possesses a peculiar nature which
defines a class of its own. The reason is partly
due to the existence of its connection with 
physical spacetime described by K\"{a}hler
geometry, which is the central core in the analytic
study of algebraic geometry. The N=2 conformal
algebra, denoted by
${\sf SCA}$, consists of the Virasoro field
$L(z)$, two super-current $G^\pm(z)$ and a
$U(1)$-current
$J(z)$, whose coefficients $L_m, J_n, G^{\pm}_p, 
(m, n \in \ZZ, p \in \frac{1}{2}+\ZZ )$, form a
super-Lie algebra with a central element $c$ : 
$$
\begin{array}{ll}
[ L_m , L_n ] = (m-n) L_{m+n} + 
\frac{c(m^3-m)}{12} \delta_{m,-n} \ , &
[ J_m , J_n ] = \frac{c m}{3}  \delta_{m, -n} \ , \
\ \ \   [L_m , J_n ] = - n J_{m+n} \ , \\

[ L_m, G_p^{\pm} ] = ( \frac{m}{2}- p ) G_{m+p}^{\pm} \ , & 
[ J_m, G_p^{\pm} ] = \pm G_{m+p}^{\pm} \ , \\
\{ G_p^+, G_q^+ \} = \{ G_p^-, G_q^- \} = 0 \ , & 
\{ G_p^+, G_q^- \} = 2 L_{p+q} + ( p-q)J_{p+q} + 
\frac{c}{3}(p^2 - \frac{1}{4}) \delta_{p, -q}  
\ .
\end{array}
$$
The algebra has a standard Cartan decomposition with
its Cartan subalgebra spanned by $c, L_0, J_0$.
An irreducible highest weight module (HWM) is
characterized by the central charge  $c$, and the
quantum numbers $h, Q$ for $L_0, J_0$
respectively, by which one defines the character
of a HWM as the following Laurent series, 
$$
{\rm NS} (z, \tau) = {\rm Tr} (
y^Q q^{h -\frac{c}{24}} ) \ , \ \ \ \ y:=
e^{2\pi\sqrt{-1}z} \ , \ q:= e^{2 \pi
\sqrt{-1}\tau} \ .
$$
The discrete series of the
unitary irreducible HWMs has the central charge
$c <3$, parametrized by
$$
c= \frac{3k}{k+2} \ , \ \ \ h=
\frac{l^2+2l-m^2}{k+2}
\ , \ \ 
\ Q= \frac{m}{k+2} \ ,
$$ 
where $1 \leq k \in \ZZ$,  $ l, m \in \ZZ$ with
$l-m \in 2\ZZ , |m| \leq l \leq k$. In rational
conformal field theory, the
representations  in the discrete series form the
building blocks of  the Gepner's models 
of string compactification \cite{Gep},
which gives rise to manifolds with
${\rm SU}(n)$ holonomy. The topological
invariants, e.g., Euler numbers, Hodge numbers and
elliptic genus, of such manifolds can be expressed
by characters of HWMs involved, (for this aspect,
see e.g.
\cite{EOTY, R92m} and references therein). Amid the
many features special for the N=2 conformal
algebra, the modular invariant property related to
the theory has drawn the constant attention from
geometers and mathematical physicists because of
its involvement with the  theory of classical
elliptic modular form. As the algebra ${\sf SCA}$
has one parameter family of equivalent generators,
which form the twisting currents, $L^a(z),
J^a(z),G^{\pm a}(z)$, one can also consider the
"twisted" character,
$$
{\rm Ch}^{(a,b)} (z, \tau) = {\rm Tr} (q^{L_0^a-\frac{{\bf c}}
{8}} (e^{2\pi \sqrt{-1} b}y)^{J_0^a} ) \ , \ \
\ \frac{-1}{2} \leq a, b \leq
\frac{1}{2} \ .
$$
The ${\rm NS}(z, \tau)$ is corresponding to $(a,
b)=(0,0)$, called the
Neveu-Schwarz sector. The Ramon sector is the one
for  $(a, b)= (\frac{-1}{2}, \frac{-1}{2})$. In a
recent work \cite{R99}, we have obtained an 
expression of the HWM characters in
the discrete series in terms of Jacobi-theta form, 
$$
{\rm Ch}_{l,m}^{(a, b)}(z, \tau) = \frac{ 
e^{ \pi\sqrt{-1}(\frac{1}{2}- \frac{l+1}{k+2})}
 \Theta^{
(\frac{1}{2} + \frac{l+1}{k+2} , \frac{1}{2})} (0,
(k+2) \tau) 
\Theta^{(a, b)} (z, \tau) }
{
\Theta^{( \frac{1}{2} +\frac{l-m+1+2a}{2(k+2)}, b )}
 ( z  , (k+2) \tau) 
\Theta^{(\frac{1}{2} - \frac{l+m+1-2a}{2(k+2)}, b
)}  ( z , (k+2) \tau) } \ 
$$
where $\Theta^{(s,t)}(z, \tau)$ is the
following expression given by the elliptic theta
function
$\vartheta^{s,t)} (z,
\tau)$ and the Dedekind
eta function $\eta ( \tau)$:
$$
\Theta^{(s,t)}  (z, \tau)  = \frac{
\vartheta^{(s,t)} (z,
\tau)}{\eta (\tau)^3 } \ , \ \ \ \ \ s, t \in \RZ \
. 
$$
Note that for the Ramon sector, the characters are
involved with the Jacobi-theta form (of weight $-1$
and index $\frac{1}{2}$). With the help of modular
form theory, we clarify the symmetries among HWMs
in the discrete series for a fixed central charge,
then obtain the Heisenberg and modular invariant
property of these characters in an explicit manner.
As an ongoing  program, with the knowledge 
obtained in N=2  conformal field theory, we shall
apply it to the geometrical study of 
$c_1=0$ manifolds as an alternative way to
understand the topological invariants of such
manifolds, especially on the elliptic genus, of
which the mathematical structure has now become
clearer after the work of Hirzebruch school
\cite{HBJ}. On the other hand, the fundamental vital
role of modular  group  has been a well-known fact 
in different areas of mathematics;
the importance of modular data has also arisen in
many physical theories other that conformal field
theory, e.g. quantum Hall effect. For the reason
that is apparent from the modular invariant 
property,  another problem   would be the
connection between   N=2 conformal
algebra and  affine algebras from the
perspective of representation theory.  It is known
that  different theories can realize the same
modular data, even within the content of conformal
field theories only. Inside
${\sf SCA}$ there is a finite dimensional super-Lie
algebra $sl (2|1)$, which  
 is geometrically realized as 
super-vector fields of the one-dimensional
projective superspace
$\PZ^{1,2}$ with 2-component Grassmann variables. 
Meanwhile characters of HWM in the discrete series
of N=2 conformal theory have shown a 
remarkable similarity to those of admissible
highest weight representations of the affine algebra
$\widehat{sl}(2|1)$ in 
\cite{KRW}. 
The highlight of our present analysis is that first,
the characters of two sides  are found to match in a
perfect way, and second, both algebras share 
analogous features on the structure. A deeper
connection is now under our investigation, and
partial results are promising. 

\section{Orbifold, Finite Group Representation
 and String Theory }
There has been spectacular progress in the
development of string theory
since its inception thirty years ago. Development
in this area has never been impeded by the lack of
experimental confirmation. Indeed, numerous bold
and imaginative strides have been taken and the
sheer elegance and logical consistency by the
arguments have served as a primary motivation for
string theorists to push their information ahead.
Meanwhile,
in the process of developing the
theory, other than employing powerful
mathematical machinery, string theorists have
proposed many elegant mathematical formulae and
conjectures, which  had never been anticipated by
mathematicians before. However these predictions
 were explained and justified only
at the level for the  need of the physical theory,
or through the physical intuition.
Yet even when definite predictions exist, an
explanation is an explanation in the physical sense,
still not a mathematical proof. With this attitude,
a rigorous mathematical justification would be
desired for further understanding the structure
so that 
results obtained would be expected to make
feedback to the original theory in physics.
Amid the many such examples occurring in the past
few decades,  a notable one would be the Vafa's
formula
 on string vacuum of the orbifoldized
Laudau-Ginzburg models in 
N=2 superconformal theory
\cite{V}. Subsequently, by using known results and
techniques in topology and algebraic geometry, the
formula which Vafa proposed  was 
justified mathematically in
\cite{R90} as the expected form of the Euler number,
Hodge number of a crepant resolution of Calabi-Yau
(possibly singular) hypersurface in a weighted
projective 4-space. Relying on this formula, 
striking numerical evidences were produced in the
early 90s, subsequently stemmed the quest of
Calabi-Yau mirror symmetry
\cite{Ya}. In the remaining section, I
shall focus on the  continuum of  such a
interdisciplinary work with progress 
recently made on the orbifold theory. 

By an orbifold, we mean an algebraic
(or analytic) variety with at most (finite group)
quotient singularities. This geometrical object was
introduced in the 1950s by
I. Satake, which he called a V-manifold, 
on his course of study of the
Siegel's modular forms
\cite{Sat}. It is a simple and natural
construction, however provides a powerful tool in
the study of various mathematical areas, including
algebraic geometry, topology, infinite dimensional
algebra representation, etc. As the procedure of an
orbifold is viewed as an  identification of certain
"known" data by a discrete twist group, one would
expect the orbifold construction will eventually 
play an equally important role in physics. Indeed
it has been shown to be the case in various physical
theories, e.g. conformal field theory, quantum
Hall effect and  string theory.  From the
geometrical aspect, the essence of an orbifold has
always been related to the study of singularity
arising from the construction.   In the mid-80s,
string theorists studied a special type of 
physical model built upon an orbifold $ M/G$ where 
$M$ is a manifold quotiented by a finite group  
$G$.  It required the following
description of the string 
vacuum in terms of the
topology of target space  \cite{DHVW}, which
was called the orbifold Euler characteristic in
\cite{HH},
\[
\chi(M, G) = \frac{1}{|G|} \sum_{
g, h \in G, \ gh=hg }\chi(M^{g, h}) \  
\]
where $M^{g, h}$
is the simultaneous fixed-point set of $g, h$. 
For the local situation where $G$ acts linearly on
$\CZ^n$,  one has the following equivalent
formulation in terms of representations of the
group $G$, 
\be
\chi(\CZ^n , G) = \# \{ {\rm irreducible \
representations \ of } \ G \} \ , \ \ \  G \subset
{\rm GL}_n(\CZ) \ .
\ele(GIrr)
For the special interest in string theory,  
$M$ is a complex manifold with a $G$-invariant
holomorphic volume form,  equivalently to say,
locally the group $G$ in (\req(GIrr)) is a subgroup
of ${\rm SL}_n(\CZ)$. So the 
$n$-dimensional orbifold
$X (= M/G)$  has at most Gorenstein singularities.
For the consistency of physical theory,  it became
immediately desirable to demand 
a crepant resolution
$\widehat{X}$ of $X$ with the compatible Euler
number: $\chi(\widehat{X} ) = \chi(M, G)$. In the
local version, it states that there exists a
resolution $\widehat{\CZ^n/G}$  of $\CZ^n/G$ for a
finite group $G$ in ${\rm SL}_n(\CZ)$, with the
canonical bundle of $\widehat{\CZ^n/G}$ trivial and
$\chi(\widehat{\CZ^n/G}) =\chi(\CZ^n, G)$ .
For
$n=2$, these groups were classified into the
renowned
$A$-$D$-$E$ series by F. Klein in 1872 (or so) in
his work of invariant theory of regular solids in
$\RZ^3$ \cite{Kl}.   The minimal resolution of a
surface singularity provides the correct solution
$\widehat{X}$
\cite{HH}. However it has long been  understood 
that the nature of "minimality" of  singularities of
algebraic (or analytical) varieties differs 
significantly as  the dimension of the
variety is larger than two. A notable example
would be the
minimal model program carried out by S. Mori and
others on the 3-dimensional birational geometry, 
which, as the prime
achievement in algebraic geometry of the 1980s, 
has provided an effective tool for the study of
algebraic 3-folds, (see \cite{Mo} and references
therein). However, the
crepant resolution of Gorenstein 3-dimensional
orbifolds was still missing in the theory. 
For the existence of crepant resolution of
3-dimensional oridifolds stemming from string
theory,  the complete mathematical solution of the
conjecture 
 was obtained in the mid-90s (see \cite{Rtop}
and references therein); the proof was based upon
techniques in toric geometry and singularity
theory, but heavily depending on the classical work
of Miller-Blichfeldt-Dickson classification of
finite groups in ${\rm SL}_3(\CZ)$ \cite{MBD} and
invariant theory of two simple groups, $I_{60}$
(icosahedral group), $H_{168}$ (Klein group)
\cite{Kl}.   The result implies that "terminal"
singularities   of 3-dimensional Gorenstein
orbifolds are in fact "smooth" in the minimal model
theory of 3-folds. Nevertheless,  due to the
computational nature of methods in the proof, the
qualitative understanding on the these crepant
resolutions was still lacking on certain
aspects from a mathematical viewpoint. 
Recently the $G$-Hilbert scheme ${\rm
Hilb}^G(\CZ^n)$,   developed by I. Nakamura $et \
al.$ \cite{Na}, has provided a conceptual approach
and understanding of crepant resolutions of
Gorenstein 3-orbifolds. Along this line, a
plausible method has been proposed on problems of
resolutions of higher dimensional orbifolds by
engaging the representation theory of finite
groups. Even for the abelian group $G$ in 
the dimension
$n=3$, the conclusion on the trivial canonical
bundle of ${\rm Hilb}^G(\CZ^3)$ would lead to a
mathematical subtlety that could not be ignored
by both physicists and mathematicians working on 
the mirror symmetry problem of Calabi-Yau 3-folds in
string theory.  A standard well known example in
Calabi-Yau mirrors is the Fermat quintic in
$\PZ^4$ with the special marginal deformation
family:
$$
X : \ \ \sum_{j=1}^5 z_i^5 + \lambda
z_1z_2z_3z_4z_5 
= 0 \ .
$$
With the maximal diagonal group $SD$ of
$z_i$s preserving the family $X$, the mirror
$X^*$ is constructed by "the" crepant resolution
of $X/SD$, $X^*= \widehat{X/SD}$  
\cite{GPR}, by which the roles of $H^{1,1},
H^{2,1}$ are interchangeable in the "quantum"
sense. On the
one-dimensional space $H^{1,1}(X)
\sim H^{2,1}(X^*)$, the choice of
crepant resolution  $\widehat{X/SD}$ makes no
difference on the conclusion. While working on 
$H^{2,1}(X)$ and $
H^{1,1}(X^*)$, many topological
invariants like Euler characteristic, elliptic
genus, have been known indifferent to the choice of
crepant resolutions, hence the same for $X^*$.
However, the topological triple intersection of
cohomologies differs for two crepant resolutions
(see, e.g.,
\cite{R93}), hence the  choice of
crepant resolution as the mirror 
$X^*=\widehat{X/SD}$ leads to the different
effect on the topological cubic form of
$H^{1,1}(X^*)$, which as the "classical" level,
the quantum triple  product of the physical
theory will be  built upon (see, e.g., articles in
\cite{Ya}). The question of the
"good" model for $X^*$ has rarely
been raised in the past, partly due to the lack of
mathematical knowledge on the issue. However, 
with the
$G$-Hilbert scheme now as the mirror $X^*$, it
 seems to have left some
fundamental  open problems on its formalism of
mirror Calabi-Yau spaces and the
question of the arbitrariness of the choice
of crepant resolutions remains a mathematical
question to be answered. 
 For $n \geq 4$,
it is known that 
a quotient orbifold $\CZ^n/G$ does not possess any
crepant resolution in general, even
for a cyclic group $G$ in the case $n=4$, (for a
selection of examples, see \cite{R97}). For
simplicity, in the discussion of  $G$-Hilbert
scheme, the quotient singularities we shall
only concern in higher dimensions are of
hypersurface type, among which certain simple group
quotient singularities are included.  For the
abelian case, the group of the $A$-type hypersurface
quotient singularity  of dimension $n$ is defined by
\begin{eqnarray*}
A_r(n) := \{ g \in {\rm SL}_n({\bf C}) \ | \ g:
\mbox{ diagonal} \ ,\  g^{r+1}=1  \} , \ \ \ \ r
\geq 1
\ ,
\end{eqnarray*}
and the orbifold can be realized as the following
hypersurface in $\CZ^{n+1}$,
$$
x^{r+1} = \prod_{j=1}^n y_j  \ , \ \ \ \ 
\ (x, y_1,
\cdots, y_n ) \in
\CZ^{n+1} \ .
$$
One can use the toric techniques 
to  study the orbifold geometry for $A_r(n)$. For
$n=4$,  we obtained the structure of
${\rm Hilb}^{A_r(4)}(\CZ^4)$ in
\cite{CR}, which is a smooth
toric variety with the non-trivial canonical bundle,
$$
\omega  = {\cal O} ( 
\sum_{k=1}^m E_k ) \ , \ \ \ m =
\frac{r(r+1)(r+2)}{6} \ ,  
$$ 
where $E_k$s are disjoint smooth
exceptional divisors in
${\rm Hilb}^{A_r(4)}(\CZ^4)$ , each of them 
is isomorphic to 
 $\PZ^1 \times \PZ^1 \times \PZ^1$ whose normal
bundle restricting to each factor $\PZ^1$ is the
$(-1)$-hyperplane bundle. Hence one can blow down
a family of
$\PZ^1$s of $E_k$ along one of the
three projections of the triple product of
$\PZ^1$. By this means, one obtains different
crepant resolutions of $\CZ^4/A_r(4)$. The relation
between those different crepant
resolutions is interpreted as 
the "flop" of 4-folds. Note that in Mori's minimal
model program of birational geometry of 3-folds,
one of the important concept introduced is the flip
operation.  Amid flip operations in the 3-fold
theory, an invertible one, called flop, was known
since long among geometers; it had played a vital
role in the study of degeneration of K3 surfaces
(see e.g. \cite{R74}). In our searching
 crepant resolutions of $A$-type hypersurface
singularity through the
$G$-Hilbert scheme, the flop operation of
dimension 4 appeared again naturally in
the process. The construction would be expected by
the common nature of varieties with the trivial
canonical bundle. The explicit description of 
 the "flop" of folds is 
defined as the connection between three
different "small"\footnote{Here the
"smallness" for a resolution  means one with the
exceptional locus of codimension $\geq 2$.}  resolutions of the isolated
singularity of an four-dimensional affine variety
$S$ with the following realization in
$\CZ^8$, 
\begin{eqnarray*}
S = \{ (x_i, y_i)_{1
\leq i \leq 4} \in \CZ^8 \ | \ \ x_iy_i= x_j y_j ,
\  x_ix_j = y_{i'}y_{j'} \ , \ 
1 \leq i \neq j \leq 4 \ \}  
\end{eqnarray*}
where the indices $i', j'$ denotes  the
complimentary pair of $i, j$. All these three
small resolutions are 4-folds with the trivial
canonical bundle. In fact, 
all the three small ones are dominated by a
smooth variety $\widetilde{S}$ over the
singular variety $S$ such that the exceptional
locus of $\widetilde{S}$ over $S$ is a divisor
isomorphic to the triple product of $\PZ^1$. Those
small crepant resolutions  are the smooth
varieties obtained by blowing-down the family 
$\PZ^1$s along one of three projections. The toric
description of these relations can be found in
\cite{LR}. On the other hand, elements of the
$G$-Hilbert scheme  parametrize certain
$G$-regular representations, which are easy to
obtain through the data of toric variety for the
case $G=A_r(4)$. Nevertheless, the analysis on
these $G$-representations displays a common
pattern for all $r$. For a simple group $G$, the
understanding of
${\rm Hilb}^G(\CZ^n)$ in terms of
$G$-representations could be an interesting problem
from both the geometrical and representation
theoretical point of view. Several examples are now
under  investigation along this scheme of
interpretation; the programme is now under
progress.

\section{Conclusion Remarks}
In this article,    
a comprehensive analysis  
and the direction of our present-day research are 
briefly reviewed on certain topics related to 
algebro-geometry approach to solvable
theories in physics. 
Though the rudimentary knowledge of a complete
mathematical theory is not yet known,  partly for
that reason most of the topics are explained
through examples, we hope the presentation would
help to give a feeling of
 how the current research is oriented.  Even in
such a formulation, many problems which 
emerged with an empirical ground  have still been 
a difficult mathematical task for the solution, and
remain to be completely understood on their
fundamental nature. However, the growing important
role of algebraic geometry in  certain
physical problems would be in evidence through our
discussion  on recent achievement and newly
progress in 2-dimensional  statistical physics and
string theory. Hopely, this consideration
will lead to impute  new insights to certain 
unsolved problems of  common interest in 
algebraic geometry and solvable quantum physics.
Meanwhile, the development of computer science in
the last decade has made possible the use of
numerical data to extract correct information   in
a wide variety of problems, both in mathematics and
physics. The power of its calculation has served us
well and continues to act as the underlying driving
force  of our  search of a theoretical conceptual
framework at a whole new level of complexity. These
experimental investigations led us to foresee the
vital role of quantitative research as a 
significant part in advancing the mathematical
knowledge  for  decades to come. In comparison to
the development of physics, it is of  interest to
note  a short remark describing the career of 1913
Nobel Laureate in physics, H. K. Onnes, related to 
his version on experimental physics around the turn
of centuries from 19th to 20th: 

\begin{em}
In 1881 he published a paper,
"Algemeene theorie de vloeistoffen" (General theory
of liquids), which dealt with the kinetic theory of
liquid state, approaching Van der Waals' law of
corresponding states from a mechanistic point of
view. This work can be considered as the beginning
of his life-long investigations into the properties
of matter at low temperatures. In his inaugural
address, " De beteekenis van het quantitatief
onderzoek in de natnurkunde" (The importance of
quantitative research in physics), he arrived at
his well-known motto "Door meten tot weten"
(knowledge through measurement), an appreciation of
the value of measurements which concerned him
through his scientific career.
\end{em}
\par \noindent
(This passage is quoted
from  "Biography of Heike Kamerlingh  Onnes, Nobel
Lectures, Physics 1901-1921", Elsevier Publishing
Company, Amsterdam.) 
H. K. Onnes' valuing on quantitative research could
serve one primary factor behind his laboratory
success  of discovering superconductivity; it would
also provide a permanent symbolic reminder as a
scientific discipline in physics as
well. In the science of mathematics, "knowledge
through computation" could be the one
philosophically more in tune with H. K. Onnes' view
on the quantitative research. Along this line at
the dawn at the new millennium, the great
challenge of how to embrace  the spectacular
computing capability provided by the successful
computer technology  in   advancing our mathematical
knowledge beneficial to physical problems will
confront  algebraic geometers in the years to come.
In concert with  this principle in the context of
solvable models, the quest of mathematical insights
and rigorous theorems to the physical theory  will
still deserve a great deal of merits from the
mathematical point of view.

\par \vspace{1cm} \noindent
{\Large\bf Acknowledgements} \par \vspace{.3cm}
\noindent  
I would like to thank  L.
Chiang, E. Date, G. von Gehlen, V. G. Kac, S. S.
Lin and M. Wakimoto for collaboration of the
research  presented in this article.
I also acknowledge the support by research programs
of the National Science Council of Taiwan over the
years when most of this work was carried out.

\end{document}